\documentstyle[12pt]{article}
\advance \textheight by \topskip
\makeatletter
\def\eqnarray{\stepcounter{equation}\let\@currentlabel=\theequation
\global\@eqnswtrue
\global\@eqcnt\z@\tabskip\@centering\let\\=\@eqncr
$$\halign to \displaywidth\bgroup\@eqnsel\hskip\@centering
  $\displaystyle\tabskip\z@{##}$&\global\@eqcnt\@ne
  \hfil$\displaystyle{\hbox{}##\hbox{}}$\hfil
  &\global\@eqcnt\tw@ $\displaystyle\tabskip\z@
  {##}$\hfil\tabskip\@centering&\llap{##}\tabskip\z@\cr}
\@addtoreset{equation}{section}
  \def\theequation{\thesection.\arabic{equation}}
\makeatother

\begin{document}

\begin{titlepage}
\hbox to \hsize{\hfil hep-th/9305133}
\hbox to \hsize{\hfil IHEP 93--48}
\hbox to \hsize{\hfil March, 1993}
\vfill
\large \bf
\begin{center}
Boundary values as Hamiltonian variables. I. \\
New Poisson brackets
\end{center}
\vskip 1cm
\normalsize
\begin{center}
{\bf Vladimir O. Soloviev\footnote{E--Mail:
vosoloviev@mx.ihep.su}}\\
{\small Institute for High Energy Physics,\\
 142284 Protvino, Moscow Region, Russia}
\end{center}
\vskip 2.cm
\begin{abstract}
\noindent
The ordinary Poisson brackets in field theory do not fulfil the Jacobi
identity if boundary values are not reasonably fixed by special boundary
conditions. We show that these brackets can be modified by adding some
surface terms to lift this restriction.  The new brackets generalize a
canonical bracket considered by Lewis, Marsden, Montgomery and Ratiu for
the free boundary problem in hydrodynamics. Our definition of Poisson
brackets permits to treat boundary values of a field on equal footing with
its internal values and directly estimate the brackets between both surface
and volume integrals.  This construction is applied to any local form of
Poisson brackets. A prescription for $\delta$-function on closed domains and
a definition of the {\it full} variational derivative are proposed.
\end{abstract}
\vfill
\end{titlepage}
\section {Introduction}

Field theory canonical formalism has some specific features which are
absent in mechanics. The need to deal with quantities
integrated over some region of space and to integrate by parts leeds in some
cases to the appearance of surface integrals in the hamiltonian and(or) in
Poisson brackets.  Mathematicians usually prefer to consider all these
surface integrals to be  zero and construct their refined formal variational
calculus by identifying integrands which are different by divergencies
\cite{gd}.  But
in field theory in some cases these surface terms are not zero and bear
physical meaning. So, this paper is devoted to an extension of the hamiltonian
formalism onto these divergent terms or informalizing  the formal
variational calculus.

It seems quite natural to require that for closed systems boundary values
should have equal rights with the internal  ones and be determined
exclusively by initial conditions together with dynamical equations. Here we
intend to decline any nondynamical boundary conditions, at least
at the first stage of investigation. Afterwards they can be restored
as any other constraints which can be put on the initial values
of dynamical variables. We believe that such an approach can help us
in the following papers of this series
to solve some physical problems that are intractable by other methods.
So, though the problem treated in this paper is a mathematical one,
our motivations are physical.

Let us  first remind some important previous results on the problem.

In the famous work by Regge and Teitelboim \cite{rt}
it was shown that to make the
hamiltonian dynamics of the gravitational field
in asymptotically flat space existent and
nontrivial it was necessary to include into the hamiltonian the
surface integrals of special kind. This paper also contains, but  not in a
very explicit form, the acknowledgement of the physical meaning
of surface integrals arising in the evaluation of Poisson brackets,
because it was found that they were in correspondence with the surface
terms in the hamiltonian through the algebra
\[
\{H(N,N^i),H(M,M^j)\}=H(L,L^k),
\]
where
\[
L=N^iM_{,i}-M^iN_{,i},
\]
\[
L^k=\gamma^{kl}(NM_{,l}-MN_{,l})+N^lM^k_{,l}-M^lN^k_{,l},
\]
and popular boundary conditions are adopted for canonical variables
$\gamma_{ij},\pi^{ij}$ and functions $N(x), M(x), N^i(x), M^j(x)$ .
In our paper \cite{s1} it was shown that this
correspondence could be exploited  under more general boundary
conditions for explicit evaluation of the surface terms in  hamiltonian.
The essence of the method was  the independence of formally defined
canonical Poisson brackets
\[
\{F,G\}=\int \Biggl[ {{\delta F}\over {\delta\gamma_{ij}(x)}}
{{\delta G}\over {\delta\pi^{ij}(x)}}-
{{\delta G}\over {\delta\gamma_{ij}(x)}}
{{\delta F}\over {\delta\pi^{ij}(x)}}\Biggr] d^3 x,
\]
on the surface integrals in the hamiltonians $F$ and $G$.
This property of
standard Poisson brackets is in full analogy with the one mentioned in Arnold's
book on classical mechanics \cite{arn}, where the two functions, defined only
up to constants, give their Poisson bracket exactly, not up to a constant.
 So, if in
mechanics the kernel of Poisson bracket consists of constants,  in field
theory the kernel of the ordinary bracket includes surface terms also.

In the papers \cite{app}, devoted to the study of the
Korteweg-de Vries(KdV)
equation, such   nonstandard features as the noncommutativity of
variational derivatives and violation of the Jacobi identity were
observed. In connection with this difficulties  different
modifications of the Gardner  bracket were proposed
\cite{app},\cite{ft},\cite{bft}.  Their comparison can be found
in Ref.~\cite{kum}. Unfortunately, these papers were unknown to us during our
work on Ashtekar's variables \cite{s2}, where  similar observations
were made.  Seemingly, such observations were made by mathematicians long
before \cite{and},\cite{ald}.  In our paper \cite{s2} it was conjectured
that the general criterion for the choice of boundary conditions in the
hamiltonian approach to field theory
should be the fulfilment of the Jacobi identity for the
standard Poisson bracket.

As we recognized from the very readable book by Olver \cite{olv},
when studying  surface waves of the ideal fluid
Lewis, Marsden, Montgomery and Ratiu (LMMR) \cite{lmmr}
proposed a modified form of canonical Poisson bracket\footnote{
It is interesting to note that the LMMR work was reported at the same
conference and published in the same journal issue as the work by Buslaev,
Faddeev and Takhtajan \cite{bft}, in which the Gardner KdV bracket had been
modified.}
\[
 \{F,G\}=\int_{{\Omega}} \Biggl[ {{\delta^{\wedge} F}\over {\delta
q(x)}}{{\delta^{\wedge} G} \over {\delta p(x)}}- {{\delta^{\wedge} G}\over
{\delta q(x)}}{{\delta^{\wedge} F}\over {\delta p(x)}} \Biggr] d^nx
\]
\[
+\oint_{\partial {\Omega}} \Biggl[ {{\delta^{\wedge} F}\over {\delta
q(x)}}\Bigg\vert _{\partial {\Omega}}{{\delta^{\vee} G} \over {\delta p(x)}}+
 {{\delta^{\vee}
F}\over {\delta q(x)}}{{\delta^{\wedge} G} \over
 {\delta p(x)}}\Bigg\vert _{\partial
{\Omega}}\Biggr] dS
\]
\[
 -\oint_{\partial {\Omega}} \Biggl[ {{\delta^{\wedge}
G}\over {\delta q(x)}}\Bigg\vert _{\partial {\Omega}}{{\delta^{\vee} F}
\over {\delta p(x)}}+
{{\delta^{\vee} G}\over {\delta q(x)}}{{\delta^{\wedge} F}
\over {\delta p(x)}}\Bigg\vert _{\partial {\Omega}}\Biggr] dS,
\]
where the two components of variational derivative were defined through
the formula:
\begin{equation}
D_{q}F(q,p)\cdot \delta q=\int_{{\Omega}}{{\delta^{\wedge}F}\over{\delta q}}
\cdot
\delta q d^nx +\oint_{\partial {\Omega}}
 {{\delta^{\vee} F}\over {\delta q}}\cdot
\delta q \vert _{\partial {\Omega}} dS,\label{eq:lmmr1}
\end{equation}
where partial Fr\'echet derivative was used in the l.h.s.,
and through the analogous formula for the momentum derivative.
 This bracket was
accompanied by  boundary conditions that LMMR considered as necessary
\begin{equation}
{{\delta^{\vee} F}\over{\delta q}}{{\delta^{\vee} G}\over{\delta p}}-
{{\delta^{\vee} G}\over{\delta q}}{{\delta^{\vee} F}\over{\delta p}}
=0.           \label{eq:lmmr2}
\end{equation}
LMMR mentioned, that  earlier the hamiltonian structure
for surface waves of the ideal
fluid for the potential flow was discovered by Zakharov \cite{zakh}.

Here we will generalize the LMMR formula
in such a way that the new bracket will fulfil the Jacobi identity without any
boundary conditions. It will permit us to consider on formally equal grounds
both volume and surface hamiltonians. The boundary values of hamiltonian
variables are now obeying their own hamiltonian equations and fixation of some
boundary conditions is simply a new constraint, that should be examined,
according to the Dirac procedure, for the presence of secondary and higher
constraints.  Such a generalization of the LMMR bracket seems necessary also
because this bracket of two differentiable functionals can be a functional not
differentiable in the sense of (\ref{eq:lmmr1}) and (\ref{eq:lmmr2}).
 Therefore, in general,
 it is not possible even to check the Jacobi identity for the LMMR bracket.

We announce more general new formulae for Poisson brackets and prove for
some important cases that they fulfil the new definition of Poisson
brackets {\it without discarding any surface integrals}.
 The cases where the
proofs are demonstrated are 1)ultralocal bracket with constant
structure matrix (the canonical case);
2)ultralocal bracket, depending on field variables
but not on their derivatives (Lie-Poisson brackets are the most popular
examples); 3)nonultralocal brackets
with constant structure matrix (Gardner-Zakharov-Faddeev bracket
for KdV equation may serve as an example).

Plan of this paper is as follows.

In Sec.2 we introduce  necessary
notations and give briefly the mathematical background: definitions,
lemmas and formulae.
Sec.3 contains the primary mathematical motivation for the construction of
the new brackets in the ultralocal case: the idea that Poisson bracket
should generate the full variation of a local functional.
In Sec.4 the method for constructing general local brackets is presented.
It is based on integration by parts of the local formula and gives some
new proposals about handling   distributions in this case.
 Of course, these
calculations can be justified only ``a posteriori'' in Sec.6.
Sec.5 is devoted to a definition of the full variational derivative as a
distribution. Here we also present the most unexpected result of
the paper: a multiplication rule for derivatives of the characteristic
function. This rule permits to write the new Poisson brackets in the
form of the old ones but with the new variational derivatives.
Sec.6 contains  three different proofs of the Jacobi identity for
the new brackets. The simplest
proof is applicable only for ultralocal brackets
with constant structure matrix.
The more general proof is applicable to  ultralocal brackets
dependent on field variables.
This proof heavily relays upon Aldersley's results \cite{ald} for higher
Eulerian operators.
The proof of Jacobi identity for nonultralocal brackets with constant
coefficients is demonstrated too. We are sure  that general
proof for arbitrary local brackets along the lines of these partial proofs
can also be constructed but it is clear, that it should be rather long.
Probably, some other ways for this proof would be shorter.
In the Conclusion we give a rather  short resume,
because a detailed comparison of our approach with Regge-Teitelboim's and
LMMR's treatments of surface terms in the Hamiltonian formalism
is postponed to next papers of this series. Appendix contains a
collection of different ways of presentation of the new brackets.

We plan to discuss the related physical problems of string theory, gauge
and gravitation fields in further papers of this series.

\vspace{12pt}

\section {Notations and mathematical background}

In this paper we  use the local coordinate language and
 instead of the manifold with a boundary consider a domain $\Omega$
in  $R^n$ having a smooth boundary $\partial\Omega$.
The characteristic function of this domain is $\theta_{\Omega}
=\theta (P_{\Omega})$,
where equation $P_{\Omega}(x^1,...,x^n)=0$ defines the boundary. We
do not expect that global formulation can meet with serious difficulties.

\medskip
{\bf Definition 2.1}
{\it An integral over  a compact domain $\Omega$
of  a function  of field variables $\phi^A(x), A=1,...,p$
and their partial derivatives $D_J\phi^A$ up to some finite order
\[
F=\int_{\Omega}d^nxf(\phi_A(x),D_J\phi_A(x))
\]
is called a local functional.}
\medskip

All the functions $f$ and $\phi_A$ as well as their variations
throughout the paper are supposed  infinitely smooth, i.e. $C^{\infty}(R^n)$.
We  use the multi-index notations $J=(j_1,...,j_n)$
\[
D_J={{\partial^{|J|}}\over{\partial^{j_1}x^1...{\partial^{j_n}x^n}}},
\qquad |J|=j_1+...+j_n.
\]
Binomial coefficients for multi-indices are
\[
{J \choose K}={j_1\choose k_1}...{j_n\choose k_n},
\]
where  ordinary binomial coefficients are
$$
{j \choose k}= \cases {
j!/(k!(j-k)!) &  if $ 0\le k \le j$; \cr
0  & otherwise. \cr }
$$
As the number of sums in some formulae of this paper is considerably
more than ten, we  write only one sign of summing without displaying
the indices of summation. According to this half-Einstein rule, sum over all
repeated indices should be understood. Only in those cases, where it is
not so, we display the summation indices. Also, we do not show the limits
of summation, because they are always natural, i.e. outside them the
summand is simply zero. Such a nice property of binomial coefficients
considerably helps us in many cases of changing of the orders of summation.
There is also a temptation to remove the useless $d^nx$ in the integrals
and to write the arguments only when they can be mixed. In principle all
integrals over finite domains would be better written over all $R^n$
with the help of characteristic function, but we will use
in parallel the notions
\[
\int_{\Omega}f\qquad\mbox{or}\qquad\int\theta_{\Omega}f
\qquad\mbox{or}\qquad\int\theta (P_{\Omega})f.
\]
We denote  as ${\cal  A}$ the space of local functionals . It is very
important that this space includes functionals with integrands
depending on derivatives of arbitrary order \cite{and2}. Otherwise the Poisson
brackets could go out of ${\cal  A}$.

\medskip
{\bf Definition 2.2}
{\it A bilinear operation $\{ \cdot , \cdot\}$
 such that for any $F,G,H\in~{\cal A}$

{\rm 1)} $\{F,G\}\in {\cal A}$;

{\rm 2)} $\{F,G\}=-\{G,F\} \qquad mod\; (Div)$;

{\rm 3)} $\{\{F,G\},H\}+\{\{H,F\},G\} +\{\{G,H\},F\}=0 \qquad mod\; (Div)$;

is called  the standard field theory Poisson bracket.  }
\medskip

{\bf Definition 2.3}
{\it A bilinear operation $\{ \cdot,\cdot\}$   such that
for any $F,G,H\in~{\cal A}$

{\rm 1)} $\{F,G\}\in {\cal A}$;

{\rm 2)} $\{F,G\}=-\{G,F\}$;

{\rm 3)} $\{\{F,G\},H\}+\{\{H,F\},G\} +\{\{G,H\},F\}=0$;

 is called the new field theory Poisson bracket.}
\medskip

{\bf Definition 2.4}\cite[Definition 5.70]{olv}
{\it Higher eulerian operators $E^J_A$
are defined through the formula of full variation of local functional}
\begin{equation}
\delta F=\sum\int_{\Omega} D_J\biggl( E^J_A(f)\delta\phi_A\biggr).
\label{eq:he1}
\end{equation}

 {\bf Lemma 2.5}\cite[Statement 5.72]{olv}
 {\it Higher eulerian operators can be given by the formula}
\begin{equation}
E^J_A(f)=\sum_K (-1)^{|K|+|J|}{K\choose J}D_{K-J}{{\partial f}
\over{\partial\phi_A^{(K)}}}.\label{eq:he}
\end{equation}

Usual variational derivative (or Euler-Lagrange derivative )
is the eulerian operator of zeroth order.
Let us mention that if $J$ is not contained in $K$, then all quantities
having multi-index $(K-J)$ are zero. The sums over $J$ in
Eq.(\ref{eq:he1}) and
over $K$ in Eq.(\ref{eq:he}) are really finite because local functional can
depend only on a finite number of derivatives according to Definition 2.1.
\medskip

{\bf Lemma 2.6}\cite[Statement 5.76]{olv}
{\it The eulerian operators have a property}
\[
E^J_A(D_I f)=E_A^{J-I}(f).
\]

Just this property was the reason for the first appearance of these operators
in paper \cite{kru}.
\medskip

{\bf Lemma 2.7}\cite[Proposition 3.1]{ald}
{\it Eulerian operator of a product of two local functionals is}
\[
E^K_A(fg)=\sum_L(-1)^{|K|+|L|}{L\choose K}
\biggl( E^L_A(f)D_{L-K}g+E^L_A(g)D_{L-K}f\biggr) .
\]

{\bf Lemma 2.8}\cite[Theorem 2.1]{ald}
{\it Product of eulerian operators is}
\[
E^I_AE^J_B(f)=\sum_K(-1)^{|K|}{{J+K}\choose J}E^{I-K}_A
{{\partial f}\over{\partial\phi_B^{(J+K)}}} .
\]

{\bf Lemma 2.9}\cite[Proposition 1.1]{ald}
\[
{{\partial f} \over
{\partial\phi_A^{(J)}}}=\sum_K {K \choose J}D_{K-J}E^K_A(f).
\]

{\bf Lemma 2.10}\cite[Lemma 2.2]{ald}
\[
{\partial \over{\partial\phi_A^{(I)}}}D_Jf=\sum_K {J \choose K}D_{J-K}
{{\partial f}\over {\partial\phi_A^{(I-K)}}},
\]

Let us mention that notations in \cite{ald} are different because
multi-indices are not used there and
because the definition of eulerian operator $E^I_A$ differs by factor
$(-1)^{|I|}$.

We also need combinatorial identities
\medskip

{\bf Lemma 2.11}\cite[Lemma 1.1]{ald}
\[
\sum^j_{l =k}(-1)^l{l \choose i}
{{j-k} \choose {l-k}}=(-1)^j{k \choose
{j-i}},
\]

that can be  written as
\[
\sum^j_{l =k}
{{(-1)^l l!} \over {(l-i)!(l-k)!(j-l
)!}}=(-1)^j{{i!k!} \over {(j-i)!
(j-k)! (i+k-j)!}}.
\]

{\bf Lemma 2.12}\cite[p.616]{pbm}
\[
\sum^j_{l=0}{i \choose l}{{j-k} \choose
{j-l}}={{i+j-k} \choose {j}},
\]

that can be also written as
\[
\sum_{l=0}^j{1 \over {l! (j-l)! (l-k)!
(i-l)!}}={{(i+j-k)!} \over {i!
j! (j-k)!(i-k)!}}.
\]
\medskip

{\bf Definition 2.13}\cite[Definition 5.28]{olv}
{\it
The partial Fr\'echet derivative of a function
$f$ is a differential operator $D_{f_A}$,
defined for arbitrary $q_A$ as}
\[
D_{f_A}(q)={d \over {d\epsilon}} f(\phi_A
 +\epsilon q_A(\phi))\Bigg\vert _{\epsilon =0}.
 \]
In our case
\begin{equation}
D_{f_A}=\sum_I {{\partial f}\over{\partial\phi_A^{(I)}}}D_I.  \label{eq:fr}
\end{equation}
The Leibnitz rule is
\begin{equation}
D_J(fg)=\sum_K {J \choose K}D_KfD_{J-K}g.\label{eq:leib}
\end{equation}

\vspace{12pt}

\section{Motivation for the new brackets from the full variation formula}

As a rule, the Poisson brackets are given by the formula
\[
\{ F,G \} =\sum\int_{\Omega}\int_{\Omega} {{\delta F} \over {\delta\phi_A(x)}}
{{\delta G} \over {\delta\phi_B(y)}}\{ \phi_A(x),\phi_B(y)\},
\]
where variational derivative is believed to be
the zeroth order eulerian operator ( Euler-Lagrange derivative)
\[
{{\delta } \over {\delta\phi_A}}=E^0_A=\sum (-1)^{|J|}D_J{{\partial}
\over {\partial\phi_A^{(J)} }}
\]
and where we do not care for any  surface integrals  because all of them
are supposed to be zero.
Here we limit our attention to ultralocal Poisson
brackets.  The more general case will be treated in the next Section.

\medskip
{\bf Definition 3.1}
{\it The standard Poisson bracket is called ultralocal if}
\[
\{\phi_A(x),\phi_B(y)\}=I_{AB}\delta (x-y),
\]
{\it where the so called implectic {\rm \cite{fufo}} operator $I_{AB}$ (or
structure matrix) can depend on the field variables $\phi_A(x)$ and their
derivatives $D_K\phi_A(x)$.}
 \medskip

These  Poisson brackets, together with the local functional $H$,
called hamiltonian, generate a variation of any local functional $F$
under fixed boundary values of $\phi_A, D_J\phi_A$ according to the formula
\[
\delta_H F=\{ F,H \} =\sum\int_{\Omega} {{\delta F} \over {\delta\phi_A}}
\delta_H\phi_A,
\]
where
\begin{equation}
\delta_H\phi_A=\sum I_{AB}{{\delta H} \over {\delta\phi_B}}.
\end{equation}

The new Poisson brackets we are searching for should analogously generate
for a given hamiltonian $H$ {\it a full variation} of a local
functional $F$ in accordance with Eq.(\ref{eq:he1})
\[
\delta_H F=\{ F,H \} =\sum\int_{\Omega}D_J\biggl( E^J_A(f)
\delta_H\phi_A\biggr) ,
\]
where variations $\delta_H\phi_A$ of field variables are linearly
dependent not only on $E^0_B(h)$, but also on higher eulerian
operators (\ref{eq:he})
\[
\delta_H\phi_A=\sum I^{(K)}_{AB}E^K_B(h).
\]
Evidently, $I^{(0)}_{AB}=I_{AB}$, and other coefficients $I^{(K)}_{AB}$
will be found below. Really, they are distributions, and this aspect will be
treated in the next two Sections. Here we need them ``in weak sense'',
 i.e. as functionals defined on the standard smooth functions.
 We will show, that these coefficients can be found from the
requirement of antisymmetry of Poisson brackets, i.e.,
\[
\sum\int_{\Omega} D_J\biggl( E^J_A(f)I^{(K)}_{AB}E^K_B(h)\biggr)
=-\sum\int_{\Omega} D_J\biggl( E^J_A(h)I^{(K)}_{AB}E^K_B(f)\biggr) .
\]
Let us  consider this condition perturbatively in the order of
eulerian operators $\vert J \vert + \vert K \vert$. In the zeroth
order the antisymmetry is fulfilled due to the related property
of the standard bracket
\[
I^{(0)}_{AB}=-I^{(0)}_{BA}.
\]
In the first order we should have
\[
\sum_{A,B}\sum_{\vert K \vert =1}\int_{\Omega}E^0_A(f)I^{(K)}_{AB}E^K_B(h)+
\sum_{A,B}\sum_{\vert J \vert =1}\int_{\Omega}D_J\biggl(E^J_A(f)I^{(0)}_{AB}
E^0_B(h)\biggr)
\]
\[
=
-\sum_{A,B}\sum_{\vert K \vert =1}\int_{\Omega}E^0_A(h)I^{(K)}_{AB}E^K_B(f)-
\sum_{A,B}\sum_{\vert J \vert =1}\int_{\Omega}D_J\biggl(E^J_A(h)I^{(0)}_{AB}
E^0_B(f)\biggr) .
\]
If we regroup the terms and exploit the zeroth order antisymmetry, then
after relabeling some indices $(A \leftrightarrow B)$, $(J \leftrightarrow
K)$
the above relation can be written as
\[
\sum_{A,B}\sum_{\vert K \vert =1}\int_{\Omega}\biggl(
E^0_A(f)I^{(K)}_{AB}E^K_B(h)+E^0_A(h)I^{(K)}_{AB}E^K_B(f)\biggr)
\]
\[
=\sum_{A,B}\sum_{\vert K \vert =1}\int_{\Omega}D_K\biggl(
E^0_A(f)I^{(0)}_{AB}E^K_B(h)+E^0_A(h)I^{(0)}_{AB}E^K_B(f)\biggr) .
\]
Taking into account the linear independence of eulerian operators
we conclude that
\begin{equation}
\sum_{A,B}\sum_{|K|=1}
\int_{\Omega}E^0_A(f)I^{(K)}_{AB}E^K_B(h)=
\sum_{A,B}\sum_{|K|=1}
\int_{\Omega}D_K\biggl(
E^0_A(f)I^{(0)}_{AB}E^K_B(h)\biggr) .\label{eq:k1}
\end{equation}
 Therefore, we succeed in
determining the coefficients $I^{(K)}_{AB}$ for the
 $|K|=1$ case.

Then, let us consider the next order
\[
\sum_{A,B}\sum_{\vert J \vert =2}\int_{\Omega} D_J\biggl( E^J_A(f)I^{(0)}_{AB}
E^0_B(h)\biggr) +
\sum_{A,B}\sum_{\vert J \vert =1}\sum_{\vert K \vert =1}
\int_{\Omega} D_J\biggl( E^J_A(f)I^{(K)}_{AB}E^K_B(h)\biggr)
\]
\[
+\sum_{A,B}\sum_{\vert K \vert =2}\int_{\Omega} E^0_A(f)I^{(K)}_{AB}
E^K_B(h)=
-\sum_{A,B}\sum_{\vert J \vert =2}\int_{\Omega} D_J\biggl( E^J_A(h)I^{(0)}_{AB}
E^0_B(f)\biggr)
\]
\begin{equation}
-\sum_{A,B}\sum_{\vert J \vert =1}\sum_{\vert K \vert =1}
\int_{\Omega} D_J\biggl( E^J_A(h)I^{(K)}_{AB}E^K_B(f)\biggr) -
\sum_{A,B}\sum_{\vert K \vert =2}\int_{\Omega} E^0_A(h)I^{(K)}_{AB}
E^K_B(f) . \label{eq:k2}
\end{equation}
If we take into account the result obtained before (\ref{eq:k1}),
then  the second terms in the l.h.s. and in the r.h.s.
of  equation (\ref{eq:k2}) are mutually cancelled. Making the
same procedure as was used for the first order  we  find
\[
\sum_{|K|=2}\int_{\Omega}E^0_A(f)I^{(K)}_{AB}E^K_B(h)=
\sum_{|K|=2}\int_{\Omega}D_K\biggl(
E^0_A(f)I^{(0)}_{AB}E^K_B(h)\biggr) .
 \]
 So, it is clear that
from the only requirement of antisymmetry we, step by
step, become convinced that the Poisson bracket should be written
as
\begin{equation}
\{ F,H\} =\sum\int_{\Omega}D_{J+K}\biggl( E^J_A(f)I_{AB}
E^K_B(h)\biggr) .\label{eq:nul}
\end{equation}
Now we are  able to formulate
\medskip

{\bf Theorem 3.2}
{\it Formula {\rm (\ref{eq:nul})} gives a new Poisson bracket,
if its zeroth order ($\vert J \vert =0=\vert K \vert$) term
is a standard ultralocal Poisson bracket
 and the structure coefficients do not depend on the derivatives
of the field variables.}
\medskip

{\it Proof.}
The antisymmetry is clear from the construction.
Evidently, the bracket is a local functional and all that we are
to prove is the Jacobi identity. This proof  will be given in Sec.6,
but  first there we will give a considerably more simple proof for the
case when $I_{AB}$ are constants.
\medskip

{\it Remark.}
It is not difficult to include the case, when $I_{AB}$ also depends on
field derivatives, but it makes the proof of Jacobi identity even longer
and the conditions for standard brackets are not so transparent.

\section{Surface terms and distributions}

The standard field theory Poisson bracket \cite{olv}
\[
\{ F,G \} =\sum\int_{\Omega}\int_{\Omega} E^0_A(f(x))
E^0_B(g(y))\{ \phi_A(x),\phi_B(y)\}
\]
is a special case, which is true only under assumption that
all surface terms arising when integrating by parts are zero,
of the formula
\begin{equation}
\{ F,G \} =\sum
\int_{\Omega}\int_{\Omega}{{\partial f} \over {\partial\phi_A^{(J)}(x)}}
{{\partial g} \over {\partial\phi_B^{(K)}(y)}}\{ D_J^{(x)}
\phi_A(x),D_K^{(y)}\phi_B(y)\},\label{eq:gen}
\end{equation}
or
\[
\{ F,G \} =\sum\int_{\Omega}\int_{\Omega} D_{f_A(x)}D_{g_B(y)}
\{ \phi_A(x),\phi_B(y)\} ,
\]
where Fr\'echet derivatives (\ref{eq:fr}) are used.
\medskip

{\bf Definition 4.1}
{\it The standard field theory
Poisson bracket  is called local if }
\begin{equation}
\{ \phi_A(x),\phi_B(y)\}={1 \over 2}\sum_L \biggl( I_{AB}^L(x)D_L^{(x)}-
I_{BA}^L(y)D_L^{(y)}\biggr) \delta (x-y),\label{eq:loc}
\end{equation}
{\it where the sum is of finite range in $|L|$.}
\medskip

Usually, derivatives over only one of the two arguments are present in
the formulae like (\ref{eq:loc}), because of the widely used relations
\begin {equation}
\biggl( D_J^{(x)}-(-1)^{|J|}D_J^{(y)}\biggr) \delta (x-y) =0,\label{eq:derdel}
\end{equation}
 also accompanied by
\[
I^L_{AB}=(-1)^{|L|+1}I^L_{BA}.
\]
But if not all surface terms, arising in integration by parts, are zero,
then (\ref{eq:derdel}) are not true. This observation was made by the
 author in \cite{s2} where
it had been realized that the problem could be reduced to the definition of
integrals like
\begin{equation}
\int_{\Omega}\int_{\Omega}
f(x)g(y)D_J^{(x)}D_K^{(y)}\delta (x-y),\label{eq:int}
 \end{equation}
 for finite domain when the test functions were nonzero on its boundary.

The theory of distributions \cite{gs} considers them as defined
on the open space domains.
In the literature known to us related problems of defining
distributions on closed domains were discussed in
books \cite{malgr},\cite{oks}, but the unique answer how to define the
 integral in (\ref{eq:int}) is absent there. So, here we propose a Rule which
 is in accordance with the results of the previous Section.
\medskip

{\bf Rule 4.2}
\begin{equation}
\int_{\Omega}\int_{\Omega} f(x)g(y)D_J^{(x)}D_K^{(y)}\delta (x-y)=
\int_{\Omega} D_KfD_Jg.\label{eq:r1}
\end{equation}

This Rule is different from that proposed  in \cite{s2}, because
the rules are compatible with  the different Poisson brackets.

Taken together, Eqs.(\ref{eq:gen}) and (\ref{eq:r1}) give us a
possibility to obtain not only the previously found expression (\ref{eq:nul})
for ultralocal brackets, but also a more general result. Let us
substitute (\ref{eq:loc}) into (\ref{eq:gen})
\[ \{ F,G \} ={1
\over 2}\sum \int_{\Omega}\int_{\Omega}{{\partial f} \over
{\partial\phi_A^{(J)}(x)}} {{\partial g} \over
{\partial\phi_B^{(K)}(y)}}
\]
\[
 \times D_J^{(x)}D_K^{(y)} \Biggl(
\biggl(  I_{AB}^L(x)D_L^{(x)}- I_{BA}^L(y)D_L^{(y)}\biggr)
\delta (x-y) \Biggr) ,
\]
and exploit the Leibnitz rule (\ref{eq:leib})
\[ \{ F,G \}
={1 \over 2}\sum \int_{\Omega}\int_{\Omega}{{\partial f} \over
{\partial\phi_A^{(J)}(x)}} {{\partial g} \over
{\partial\phi_B^{(K)}(y)}}
\]
\[
 \times\biggl( {J \choose M}
D_M^{(x)} I_{AB}^L(x)D_{L+J-M}^{(x)}D_K^{(y)} -{K \choose
M}D_M^{(y)}I_{BA}^L(y)D_J^{(x)} D_{L+K-M}^{(y)}\biggr) \delta (x-y).
\]
Then take off one of the integrations by Rule 4.2
\[
\{ F,G \}
  ={1 \over 2}\sum \int_{\Omega}\Biggl( {J \choose M}D_{L+J-M}
{{\partial g} \over {\partial\phi_B^{(K)}}}D_K\biggl(
{{\partial f} \over {\partial\phi_A^{(J)}}}
D_M I_{AB}^L\biggr)
\]
\[
-{K \choose M}D_{L+K-M}
{{\partial f} \over {\partial\phi_A^{(J)}}}
D_J \biggl(
{{\partial g} \over {\partial\phi_B^{(K)}}}
D_MI_{BA}^L\biggr) \Biggr) .
 \]
Once more using the Leibnitz rule
\[
\{ F,G \} ={1 \over 2}\sum
\int_{\Omega}\Biggl( {J \choose M}{K \choose N}D_{L+J-M}
{{\partial g} \over {\partial\phi_B^{(K)}}}D_{N+M}I_{AB}^L
D_{K-N}{{\partial f} \over {\partial\phi_A^{(J)}}}
\]
\[
-{K \choose M}{J \choose N}D_{L+K-M}
{{\partial f} \over {\partial\phi_A^{(J)}}}
D_{N+M}I_{BA}^LD_{J-N}
{{\partial g} \over {\partial\phi_B^{(K)}}}
 \Biggr) ,
\]
and making  changes $(J \leftrightarrow K)$,
$(A \leftrightarrow B)$ in the second term we obtain
\[
\{ F,G \} ={1 \over 2}\sum
\int_{\Omega} {J \choose M}{K \choose N}D_{N+M}I_{AB}^L
\]
\[
\times \biggl(
D_{K-N}{{\partial f} \over {\partial\phi_A^{(J)}}}
 D_{L+J-M}{{\partial g} \over {\partial\phi_B^{(K)}}}
 - D_{K-N}{{\partial g} \over {\partial\phi_A^{(J)}}}
 D_{L+J-M}{{\partial f} \over {\partial\phi_B^{(K)}}}
 \biggr) .
\]
Transform here the partial
derivatives into eulerian operators  according to Lemma 2.9
\[
\{ F,G \} ={1 \over 2}\sum
\int_{\Omega} {J \choose M}{K \choose N}{P \choose J}{Q \choose K}
D_{N+M}I_{AB}^L
\]
\[
\times\biggl(
D_{P-J+K-N}E^P_A(f)
D_{Q-K+L+J-M}E^Q_B(g)
  - (F \leftrightarrow G) \biggr) ,
\]
make a change $J \rightarrow J+K$ and estimate the sum over $K$
according to Lemma 2.12
\[
\sum_K {{J+K} \choose M}{P \choose {J+K}}{K \choose N}{Q \choose K}
={P \choose M}{Q \choose N}{{P+Q-M-N} \choose {P-J-N}}.
\]
Then we get
\[
\{ F,G \}={1 \over 2}\sum
\int_{\Omega} {P \choose M}{Q \choose N}{{P+Q-M-N} \choose {P-J-N}}
\]
\[
\times D_{N+M}I_{AB}^L
\biggl( D_{P-J-N}E^P_A(f)
D_{Q+L+J-M}E^Q_B(g)
  - (F \leftrightarrow G) \biggr) .
\]
It is not difficult to get convinced with the help of Leibnitz rule that the
obtained result coincides with
\begin{equation}
{1 \over 2}\sum\int_{\Omega} D_{P+Q}\biggl(
E^P_A(f)I^L_{AB}D_LE^Q_B(g) - (F \leftrightarrow G) \biggr) .\label{eq:nloc}
\end{equation}
For ultralocal case $I^L_{AB}=\delta_{L0}I_{AB}$, $I_{AB}=-I_{BA}$,
and, evidently, Eq.(\ref{eq:nul}) can be reproduced.  For a
more general case we have
\medskip

{\bf Theorem 4.3}
{\it The new Poisson brackets, corresponding to the standard local
Poisson brackets with constant structure matrix, are given by
formula {\rm (\ref{eq:nloc})}. }
\medskip

{\it Proof.} The antisymmetry is evident, it is also evident
that (\ref{eq:nloc}) gives local functional.
 The Jacobi identity for this case will be proved in Sec.6.3.
\medskip

{\it Remark.}
Evidently, the construction does not restrict us to the case $I^L_{AB}=const$.
But the proof of Jacobi identity becomes  more difficult in the general case.

\section{A full variational derivative and a multiplication rule:
towards informal variational calculus}

Let us present
the standard variational, or Euler-Lagrange,  derivative in the form
\[
{{\delta F} \over {\delta\phi_A(x)}}=E^0_A(f)\theta_{\Omega}.
\]
Then it gives us a full variation
\[
\delta F=\sum\int {{\delta F} \over
{\delta\phi_A}}\delta\phi_A,
 \]
 of a local functional
\[
F=\int \theta_{\Omega}f(\phi_A, D_J\phi_A),
\]
only if all surface integrals in
the general formula
 \[
\delta F=\sum\int\theta_{\Omega}D_J\biggl(
E^J_A(f)\delta\phi_A\biggr),
 \]
 are zero.

\medskip
{\bf Definition 5.1}
{\it A distribution ${\delta F}/ {\delta\phi_A}$ such
that in general case, i.e. for arbitrary smooth variations $\delta\phi_A(x)$,
\begin{equation}
\delta F=\sum\int{{\delta F} \over{\delta\phi_A}}\delta\phi_A,
\end{equation}
will be called the full variational derivative of a local functional $F$.}
\medskip

{\bf Statement 5.2}
{\it The full variational derivative can be written in the form
\begin{equation}
{{\delta F} \over {\delta\phi_A}}=\sum (-1)^{|J|}E^J_A(f)D_J\theta_
{\Omega},\label{eq:varfull}
\end{equation}
where $\theta_{\Omega}$ is a characteristic function of the domain
of integration $\Omega$.}
\medskip

{\it Proof.}
Through integration by parts
\[
\sum\int (-1)^{|J|}E^J_A(f)D_J\theta_{\Omega}\delta\phi_A=
\sum\int \theta_{\Omega}D_J\biggl( E^J_A(f)\delta\phi_A
\biggr)
\]
\[
=\sum\int_{\Omega} D_J\biggl( E^J_A(f)\delta\phi_A
\biggr).
\]
\medskip

{\bf Statement 5.3}
{\it By using  {\rm Definition 5.1} the
new Poisson brackets {\rm (\ref{eq:nloc})} can be written in the form
\begin{equation}
\{ F,G \} =\sum\int\int
{{\delta F} \over {\delta\phi_A(x)}}\{ \phi_A(x),\phi_B(y) \}
{{\delta G} \over {\delta\phi_B(y)}},\label{eq:nbvar}
\end{equation}
if we admit the following  multiplication rule:}
\medskip

{\bf Rule 5.4}
\[
D_J\theta (P_{\Omega}) \times D_K\theta (P_{\Omega})=D_{J+K}\theta
 (P_{\Omega}).
\]

{\it Remark.} Of course, this Rule has its domain of applicability only
inside our procedure of calculating the Poisson brackets. Maybe, it can
also find its place in the new theory of generalized functions \cite{eg}.
\medskip

{\it Proof of the Statement.}
Let us substitute (\ref{eq:varfull}) and (\ref{eq:loc})
into (\ref{eq:nbvar}) and take off the
derivatives from the $\delta$-function through integration
by parts
\[
{1 \over
2}\sum (-1)^{|J|+|K|+|L|}\int\int \Biggl[ D_L\biggl(D_J(\theta
(P_{\Omega}(x))E^J_A(f)I^L_{AB}(x)\biggr) D_K\theta (P_{\Omega}(y))
E^K_B(g)
\]
\[
-D_L\biggl( D_K\theta (P_{\Omega}(y))E^K_B(g)I^L_{BA}(y)\biggr)
D_J\theta(P_{\Omega}(x))E^J_A(f)\Biggr] \delta (x-y).
\]
Then take off one of  integrations
with the help of  $\delta$-function, and afterwards use
the Leibnitz rule
\[
{1 \over 2}\sum (-1)^{|J|+|K|+|L|}{L \choose M}\int \Biggl[
D_{J+M}\theta_{\Omega}
 D_K\theta_{\Omega} E^K_B(g)D_{L-M}\biggl( E^J_A(f)I^L_{AB}\biggr)
\]
\[
-D_{K+M}\theta_{\Omega} D_J\theta_{\Omega}
 E^J_A(f)D_{L-M}\biggl(
E^K_B(g)I^L_{BA}\biggr) \Biggr] .
 \]
 After exploiting  Rule 5.4
and integrating by parts we have
 \[
{1 \over 2}\sum
(-1)^{|L|+|M|}{L \choose M}
\int_{\Omega} D_{J+K+M}\Biggl[ E^K_B(g)D_{L-M}\biggl(
E^J_A(f)I^L_{AB}\biggr)
\]
\[
 -E^J_A(f)D_{L-M}\biggl(
E^K_B(g)I^L_{BA}\biggr)\Biggr],
 \]
and after using the Leibnitz rule
 \[
{1 \over 2}\sum (-1)^{|L|+|M|}{L \choose M}{M \choose
N}\int_{\Omega}D_{J+K}\Biggl[ D_NE^K_B(g)D_{L-N}
\biggl( E^J_A(f)I^L_{AB}\biggr)
\]
\[
 -D_NE^J_A(f)D_{L-N}\biggl( E^K_B(g)I^L_{BA}\biggr)\Biggr].
\]
Summing over $M$
\begin{equation}
\sum_M(-1)^{|M|}{L \choose M}{M \choose N}=(-1)^{|L|}\delta_{L,N},
\end{equation}
completes the proof by giving Eq.(\ref{eq:nloc}).
\medskip

{\bf Statement 5.5}
{\it  {\rm Rule 4.2} is a corollary of  {\rm Rule 5.4}}.
\medskip

{\it Proof.}
With the help of characteristic function $\theta_{\Omega}$ we can write the
l.h.s. of (\ref{eq:r1}) in the form of integrals over infinite space $R^n$
\[
\int \int \theta(P_{\Omega} (x))f(x)\theta(P_{\Omega}(y))g(y)
D_J^{(x)}D_K^{(y)}\delta (x-y),
\]
then no surface terms arise after integration by parts and we have
\[
(-1)^{|J|+|K|}\int\int D_J^{(x)}\biggl( \theta (P_{\Omega}(x))
f(x)\biggr) D_K^{(y)}\biggl( \theta (P_{\Omega}(y))g(y)\biggr) \delta (x-y).
\]
Remove one of the two integrations with the help of the $\delta$ -function
and obtain
\[
(-1)^{|J|+|K|}\int  D_J(\theta_{\Omega} f)D_K(\theta_{\Omega} g),
\]
then use the Leibnitz rule
\[
(-1)^{|J|+|K|}\sum_{L,M}{J \choose L}{K \choose M}\int D_L\theta_{\Omega}
D_M\theta_{\Omega} D_{J-L}fD_{K-M}g,
\]
and  Rule 5.4. After one more integration by parts over $R^n$
we obtain an integral over $\Omega$
\[
\sum_{L,M}(-1)^{|J|+|K|+|L|+|M|}{J \choose L}{K \choose M}\int_{\Omega}
D_{L+M}\biggl( D_{J-L}fD_{K-M}g\biggr).
\]
Again exploiting the Leibnitz rule and, afterwards, calculating
the sum over $M$,
\[
\sum_M (-1)^{|M|}{K \choose M}{{L+M} \choose {N}}=(-1)^{|K|}{{L} \choose
{N-K}},
\]
we obtain
\[
\sum_{L,N} (-1)^{|J|+|L|}{J \choose L}{{L} \choose {N-K}}\int_{\Omega}
D_{N+J-L}fD_{K+L-N}g.
\]
After making a change $N \rightarrow N+L$ and calculating the sum over
$L$ according to Lemma 2.11 we get the r.h.s. of (\ref{eq:r1}). The proof
is completed.

\section{Proofs of Jacobi identity}

\subsection{A toy proof}

{\bf Statement 6.1.1}
{\it For constant structure matrix the Poisson bracket {\rm (\ref{eq:nul})}
can be written in the form}
\begin{equation} \{ F,G\}
=\sum\int_{\Omega}Tr(D_{f_A}I_{AB}D_{g_B}), \label{eq:nulc}
\end{equation}
 {\it where $D_{f_A}$
is Fr\'echet derivative {\rm (\ref{eq:fr})}, and}
\[
Tr(D_{f_A}I_{AB}D_{g_B})=\sum I_{AB}D_J{{\partial f}\over
{\partial\phi_A^{(I)}}}D_I {{\partial g}\over{\partial\phi_B^{(J)}}}.
\]

{\it Proof.}
Let us use the Leibnitz rule in Eq.(\ref{eq:nul})
\[
\{ F,G \}
=\sum I_{AB}\int_{\Omega}
{{I+J}\choose M}D_ME^I_A(f)D_{I+J-M}E^J_B(g).
\]
By exploiting  Lemma 2.5 it can be transformed to
\[
\sum I_{AB}
(-1)^{\vert I \vert +\vert K \vert +\vert J\vert+\vert L\vert}
{{I+J}\choose M}{K \choose I}{L \choose J}
 \int_{\Omega} D_{M+K-I}
{{\partial f}\over{\partial\phi_A^{(K)}}} D_{I-M+L} {{\partial
g}\over{\partial\phi_B^{(L)}}}.
 \]
 Then by changing the indices $M
\rightarrow M+I$ and the order of summation
with the help of Lemmas 2.11 and 2.12
we are able to estimate the sum
\[
\sum_{I,J} (-1)^{\vert I\vert +\vert J\vert}{{I+J}
 \choose {I+M}}{K \choose I}{L
\choose J}=
(-1)^{\vert K\vert+\vert L\vert}\delta_{M,L-K}.
\]
 As a result we obtain (\ref{eq:nulc}).
\medskip

{\bf Statement 6.1.2}
{\it The Poisson bracket, given by formula {\rm (\ref{eq:nul})}, fulfils
the Jacobi identity when $I_{AB}=const$.}
 \medskip

{\it Proof.}
Let us transform the expression
\[
\{ \{ F,G \} ,H \}=\sum I_{AB}I_{CD}\int_{\Omega}
D_I{{\partial h} \over{\partial \phi^{(J)}_B}}
\]
\[
 \times D_J
\Biggl( {{\partial} \over {\partial \phi_A^{(I)}}}(D_K{{\partial f} \over
{\partial \phi_C^{(L)}}})D_L{{\partial g} \over {\partial \phi_D^{(K)}}}+
D_L{{\partial f} \over {\partial \phi_C^{(K)}}}
{{\partial} \over {\partial \phi_A^{(I)}}}(D_K{{\partial g} \over
{\partial \phi_D^{(L)}}})\Biggr) ,
\]
with  Lemma 2.10, Leibnitz rule and antisymmetry over $C \leftrightarrow D$
\[
\{ \{ F,G \} ,H \}=\sum
{K \choose M}{J  \choose N} I_{AB}I_{CD}
\]
\[
\times \int_{\Omega}\Biggl( D_{K+N-M}{{\partial^2 f} \over {\partial
\phi_A^{(I-M)}\partial \phi^{(L)}_C}} D_I{{\partial h} \over{\partial
\phi_B^{(J)}}}D_{J+L-N} {{\partial g} \over {\partial \phi^{(K)}_D}}
\]
\[
-D_{K+N-M}{{\partial^2 g}\over {\partial \phi^{(I-M)}_A \partial \phi^{(L)}
_C}}D_I{{\partial h} \over{\partial \phi^{(J)}_B}}D_{J+L-N}
{{\partial f} \over {\partial \phi_D^{(K)}}}\Biggr) .
\]
After cyclic permutation we have
\[
\{ \{ F,G \} ,H \} +\{ \{ H,F \} ,G \} +\{ \{ G,H \} ,F \}
\]
\[
=\sum{K \choose M}{J \choose N}
 I_{AB}I_{CD}\int_{\Omega}
D_{K+N-M}{{\partial^2 f} \over
{\partial \phi_A^{(I-M)}\partial \phi^{(L)}_C}}
\]
\[
\times \Biggl( D_{J+L-N}{{\partial g} \over {\partial \phi^{(K)}_D}}
D_I{{\partial h} \over{\partial \phi^{(J)}_B}}-(g \leftrightarrow h)\Biggr)
 +\ldots,
\]
where dots mean analogous terms with cyclically permuted $f,g,h$.
By changing indices $N \rightarrow N+J$, $I \rightarrow I+M$, we get
\[
\{ \{ F,G \} ,H \} +\{ \{ H,F \} ,G \} +\{ \{ G,H \} ,F \}
=\sum{K \choose {K-M}}{J \choose {J+N}}
 I_{AB}I_{CD}
\]
\[
\times \int_{\Omega}D_{J+K+N-M}{{\partial^2 f} \over
{\partial \phi_A^{(I)}\partial \phi^{(L)}_C}}
\Biggl[ D_{L-N}{{\partial g} \over {\partial \phi^{(K)}_D}}
D_{I+M}{{\partial h} \over{\partial \phi^{(J)}_B}}-(g \leftrightarrow h)
\Biggr]
+\ldots
\]
So, if we simultaneously change $I \leftrightarrow L$, $J \leftrightarrow
K$, $M \leftrightarrow -N$, $A \leftrightarrow C$ and  $B \leftrightarrow D$,
then the expression in square brackets changes its sign while the
coefficient before the brackets
 stands as itself. Therefore, the sum equals zero, and
the Jacobi identity is fulfilled in this simplest case.

\subsection{Proof  for ultralocal case}

{\bf Statement 6.2.1}
{\it Ultralocal Poisson brackets, given by formula {\rm (\ref{eq:nul})},
with the
coefficients, depending on field variables but not on their derivatives,
exactly satisfy the Jacobi identity, if the corresponding standard brackets
satisfy it up to total divergences.}
\medskip

{\it Proof.}
Let us transform the expression
\[
\{ \{ F,G \} ,H \}=\sum \int_{\Omega}
D_{I+J}\Biggl( E^I_A\biggl( D_{K+L}\bigl( E^K_C(f)I_{CD}
E^L_D(g)\bigr)\biggr) I_{AB}E^J_B(h)\Biggr)
\]
according to Lemmas 2.6 and 2.7. Then, taking into account that
\[
E^M_A(I_{CD})=\delta_{M0}{{\partial I_{CD}} \over {\partial \phi_A}},
\]
we obtain
\[
\sum
(-1)^{\vert I \vert +\vert K \vert +\vert L \vert + \vert M \vert}
{M \choose {I-K-L}}\int_{\Omega}D_{I+J}\Biggl( E^J_B(h)I_{AB}
\biggl[\delta_{M0}
{{\partial I_{CD}} \over {\partial \phi_A}}
\]
\begin{equation}
\times D_{M+K+L-I}
\biggl( E^K_C(f)E^L_D(g)\biggr)
+D_{M+K+L-I}I_{CD}E^M_A\biggl( E^K_C(f)E^L_D(g)\biggr)\biggr] \Biggr).
\label{eq:ul1}
\end{equation}
Let us consider the first term in square brackets. As
$M=0$, then the binomial coefficient is not zero only when $I=K+L$,
therefore this term becomes
\begin{equation}
\sum\int_{\Omega}D_{J+K+L}\Biggl( I_{AB}
{{\partial I_{CD}} \over {\partial \phi_A}}E^K_C(f)E^L_D(g)E^J_B(h)\Biggr) .
\label{eq:ul2}
\end{equation}
After cyclic permutation of $F,G,H$ and having in mind the symmetry
in $J,K,L$ we see that this term gives no impact on the r.h.s.
of Jacobi identity if
\begin{equation}
I_{AB}{{\partial I_{CD}} \over {\partial \phi_A}}+
I_{AD}{{\partial I_{BC}} \over {\partial \phi_A}}+
I_{AC}{{\partial I_{DB}} \over {\partial \phi_A}}=0.\label{eq:ul3}
\end{equation}
But just this condition is necessary \cite{olv} for fulfilment of the Jacobi
identity by standard Poisson brackets in this case.
Therefore, we can now  care only for  the second term in (\ref{eq:ul1}).

Once more exploit Lemma 2.7, then the term of interest is equal to
\[
\sum
(-1)^{\vert I \vert +\vert K \vert +\vert L \vert + \vert N \vert}
{M \choose {I-K-L}}{N \choose M}\int_{\Omega}D_{I+J}\Biggl( I_{AB}E^J_B(h)
\]
\[
\times  D_{M+K+L-I}I_{CD}\biggl[
E^N_AE^K_C(f)D_{N-M}E^L_D(g)
+E^N_AE^L_D(g)D_{N-M}E^K_C(f)\biggr] \Biggr).
\]
If we take into account antisymmetry of the coefficient before the
square bracket under $C \leftrightarrow D$ and its symmetry under
$K \leftrightarrow L$, we obtain
\[
\sum
(-1)^{\vert I \vert +\vert K \vert +\vert L \vert + \vert N \vert}
{M \choose {I-K-L}}{N \choose M}\int_{\Omega} D_{I+J}\Biggl(
I_{AB}D_{M+K+L-I}I_{CD}
\]
\[
\times E^J_B(h)\biggl[
E^N_AE^K_C(f)D_{N-M}E^L_D(g)
-E^N_AE^K_C(g)D_{N-M}E^L_D(f)\biggr] \Biggr).
\]
After cyclic permutation of $F,G,H$ this expression can be written as
\[
\sum
(-1)^{\vert I \vert +\vert K \vert +\vert L \vert + \vert N \vert}
{M \choose {I-K-L}}{N \choose M}\int_{\Omega}
D_{I+J}\Biggl( E^N_AE^K_C(f)
\]
\[
\times  I_{AB}D_{M+K+L-I}I_{CD}\biggl[ E^J_B(h)
D_{N-M}E^L_D(g)-(G \leftrightarrow H)\biggr]\Biggr) +\ldots.
\]
Let us exploit the Leibnitz rule and get
\[
\sum
(-1)^{\vert I \vert +\vert K \vert +\vert L \vert + \vert N \vert}
{M \choose {I-K-L}}{N \choose M}{{I+J} \choose P}
\]
\[
\times {{I+J-P} \choose Q}
{{I+J-P-Q} \choose R}{{I+J-P-Q-R} \choose S}
\]
\[
 \times\int_{\Omega}
D_PI_{AB}D_{Q+M+K+L-I}I_{CD}
 D_RE^N_AE^K_C(f)
\]
\[
\times\Biggl[
D_{S+N-M}E^L_D(g)D_{I+J-P-Q-R-S}E^J_B(h)-
(G \leftrightarrow H))\Biggr] +\ldots.
\]
Transform the coefficient before the square bracket according to Lemmas 2.8
and 2.5, i.e., make a substitution
\[
D_RE^N_AE^K_C(f)=\sum_{T,U}(-1)^{\vert U \vert +\vert N \vert}
{{K+T} \choose K}{U \choose {N-T}}D_{R+T+U-N}{{\partial^2 f} \over
{\partial \phi^{(U)}_A \partial \phi_C^{(K+T)}}}.
\]
After that, it is possible to simplify the expression before the
square brackets by changing indices, order of summation and
explicit calculation of the four sums of binomial coefficients.

First, make the changes $T \rightarrow T-K$, $M \rightarrow M-K$,
$N \rightarrow N-K$ and estimate the sum over $K$ according to Lemma 2.11
\[
\sum_K(-1)^{\vert K \vert}{T \choose K}{{M-K} \choose {I-K-L}}
{{N-K} \choose {M-K}}={{N-T} \choose {I-L}}{{L+N-I} \choose {N-M}},
\]
(during its calculation a trivial shift of argument is exploited,
later we will not mention these details).

Then make redefinitions $Q \rightarrow Q+I-M-L$, $S \rightarrow S+M-N$,
$R \rightarrow R+N$ and calculate the sum over $M$ by exploiting
Lemma 2.12
\[
\sum_M{{I+J-P} \choose {I+Q-M-L}}{{L+N-I} \choose {N-M}}
{{J+M+L-P-Q-R-N} \choose {S+M-N}}
\]
\[
\times{{J+M+L-P-Q} \choose {R+N}}=
{{Q+S} \choose Q}{{I+J-P} \choose {N+R}}{{I+J-P-N-R}\choose {I+Q+S-N-L}}.
\]
After a new replacement $R \rightarrow J+L-P-Q-R-S$
we can estimate the sum over $N$
\[
\sum_N
{{I+J-P} \choose{N+J+L-P-Q-R-S}}{{I-N-L+Q+S-R} \choose{I+Q+S-N-L}}
\]
\[
\times {U \choose{N-T}}{{N-T} \choose{I-L}}
={{J+L+U-P-R} \choose{Q+S-T}}{U \choose{I-L}}{{I+J-P} \choose R}.
\]
And the last summation is
\[
\sum_I(-1)^{\vert I \vert}{U  \choose{I-L}}{{I+J-P} \choose R}{{I+J} \choose
P}=(-1)^{\vert U \vert +\vert L \vert}{{P+R}  \choose P}{{L+J} \choose
{P+R-U}}
\]
As a result, the term under study becomes
\[
\sum {{Q+S} \choose Q}{{P+R} \choose P}
{{L+J} \choose {P+R-U}}{{L+J+U-P-R} \choose {Q+S-T}}
\int_{\Omega}
 D_PI_{AB}
\]
\[
\times D_QI_{CD}D_{J+L+T+U-P-Q-R-S}
{{\partial^2 f} \over {\partial \phi_A^{(U)}\partial \phi_C^{(T)}}}
\biggl[ D_RE^J_B(h)D_SE^L_D(g)-(H \leftrightarrow G)\biggr] .
\]
It is not difficult to see
that under the simultaneous change $A \leftrightarrow C$, $B \leftrightarrow
D$, $J \leftrightarrow L$, $R \leftrightarrow S$, $U \leftrightarrow T$, $P
\leftrightarrow Q$ the square bracket changes
 its sign whereas the coefficient
before it does not, i.e. all the expression equals zero. With
the previous study (\ref{eq:ul2}), (\ref{eq:ul3}) in mind the proof is
completed.

\subsection{Proof for nonultralocal case}

{\bf Statement 6.3.1}
{\it Nonultralocal Poisson brackets given by  formula {\rm (\ref{eq:nloc})},
 exactly satisfy the Jacobi identity, when $I^K_{AB}=const$.}
\medskip

{\it Proof.}
By Lemma 2.6 and changes $I \leftrightarrow J$, $A \leftrightarrow B$ we have
\[
\{ \{ F,G \} ,H \} =
{1 \over 4}\sum I^N_{CD}\int_{\Omega}D_{I+J}\Biggl( \biggl(
I^K_{AB}D_KE^J_B(h)-E^J_B(h)I^K_{BA}D_K\biggr)
\]
\[
\times E^{I-L-M}_A\biggl( E^L_C(f)D_NE^M_D(g)-(F \leftrightarrow
G)\biggr)\Biggr) ,
\]
and by using Lemma 2.7  obtain
\[
E^{I-L-M}_A\biggl( E^L_C(f)D_NE^M_D(g)\biggr) =
\sum_P(-1)^{\vert P \vert +\vert I \vert +\vert L \vert +\vert M \vert}
{P \choose {I-L-M}}
\]
\[
\times \biggl( E^P_AE^L_C(f)D_{P-I+L+M+N}E^M_D(g)+E^{P-N}_AE^M_D(g)
D_{P-I+L+M}E^L_C(f)\biggr) .
\]
Then let us exploit the symmetry $L \leftrightarrow M$ and
change $C \leftrightarrow D$
\[
\{ \{ F,G \} ,H \} ={1 \over 4}
\sum(-1)^{\vert P \vert +\vert I \vert +\vert L \vert +\vert M \vert}
{P \choose {I-L-M}}
\]
\[
\times\int_{\Omega}D_{I+J}
\Biggl( \biggl(
I^K_{AB}D_KE^J_B(h) - E^J_B(h)I^K_{BA}D_K\biggr)
 \biggl( I^N_{CD}E^P_AE^L_C(f)
\]
\[
\times D_{L+M+P-I+N}E^M_D(g)-
I^N_{DC}E^{P-N}_AE^L_C(g)D_{L+M+P-I}E^M_D(f)\biggr)\Biggr).
\]
Make a change $P \rightarrow P+N$ in the second term
\[
{1 \over 4}\sum \Biggl[ {P \choose {I-L-M}}I^N_{CD}-(-1)^{\vert N \vert}
{{P+N} \choose {I-L-M}}I^N_{DC}\Biggr]
\]
\[
\times (-1)^{
\vert P \vert +\vert I \vert +\vert L \vert +\vert M \vert}
 \int_{\Omega}D_{I+J}
\Biggl( \biggl(I^K_{AB}D_KE^J_B(h)-E^J_B(h)I^K_{BA}D_K\biggr)
\]
\[
\times \biggl( E^P_AE^L_C(f)D_{M+L+P+N-I}E^M_D(g)- (F \leftrightarrow G)
\biggr)\Biggr).
\]
Then calculate $D_K$ according to the Leibnitz rule
\[
\sum_Q{K \choose Q}\biggl( D_{K-Q}E^P_AE^L_C(f)
D_{Q+M+L+P+N-I}E^M_D(g)- (F \leftrightarrow G) \biggr)
\]
and, analogously, $D_{I+J}$
\[
\{ \{ F,G \} ,H \} ={1 \over 4}\sum
\Biggl[ {P \choose {I-L-M}}I^N_{CD}-(-1)^{\vert N \vert}I^N_{DC}
{{P+N} \choose {I-L-M}}\Biggr]
\]
\[
\times(-1)^{\vert P \vert +\vert I \vert +\vert L \vert +\vert M \vert}
{{I+J} \choose R}{{I+J-R} \choose S}\int_{\Omega}
\Biggl[ I^K_{AB}D_{R+K}E^J_B(h)
\]
\[
\times\biggl( D_SE^P_AE^L_C(f)D_{J-
R-S+M+L+P+N}E^M_D(g)-(F \leftrightarrow G)\biggr) -
I^K_{BA}D_RE^J_B(h)
\]
\[
\times \sum_Q {K \choose Q}\biggl( D_{S+K-Q}E^P_AE^L_C(f)
D_{J-R-S+M+L+N+P+Q}E^M_D(g)-(F \leftrightarrow G)\biggr)\Biggr]
\]
Now we are able to sum over $I$ according to Lemma 2.11
$$
\sum_I(-1)^{|I|}{P \choose {I-L-M}}{{I+J} \choose R}
{{I+J-R} \choose S}
$$
$$
=(-1)^{|P|+|L|+|M|}{{R+S} \choose R}{{J+L+M} \choose {R+S-P}},
$$
$$
\sum_I(-1)^{|I|}{{P+N} \choose {I-L-M}}{{I+J} \choose R}
{{I+J-R} \choose S}
$$
$$
=(-1)^{|P|+|L|+|M|+|N|}{{R+S} \choose R}{{J+L+M} \choose
{R+S-P-N}},
$$
and obtain
$$
\{ \{ F,G \} ,H \} ={1 \over 4} \sum \Biggl[
{{J+L+M} \choose {R+S-P}}I^N_{CD}-I^N_{DC} {{J+L+M} \choose {R+S-P-N}}\Biggr]
{{R+S} \choose R}
$$
$$
\times \int_{\Omega}\Biggl( I^K_{AB}D_{R+K}E^J_B(h)
\biggl( D_SE^P_AE^L_C(f)D_{J-R-S+M+L+P+N}E^M_D(g)-(F \leftrightarrow G)\biggr)
$$
$$
-I^K_{BA}D_RE^J_B(h)\sum_Q{K \choose Q}
$$
$$
\times\biggl( D_{J-R-S+M+L+N+P+Q}E^M_D(g)
D_{S+K-Q}E^P_AE^L_C(f)-(F \leftrightarrow G)\biggr) \Biggr).
$$
If changes $R \rightarrow R-K$ are made in the first term
\[
{1 \over 4}\sum\Biggl[ {{J+L+M} \choose {R+S-P-K}}I^N_{CD}-I^N_{DC}
{{J+L+M} \choose {R+S-K-P-N}}\Biggr] {{R+S-K} \choose
{R-K}}
\]
\[
\times \int_{\Omega}
I^K_{AB}D_RE^J_B(h)\biggl( D_SE^P_AE^L_C(f)D_{J- R-S+M+L+P+N+K}E^M_D(g)-(F
\leftrightarrow G)\biggr) ,
\]
and $S \rightarrow S-K+Q$ in the second
\[
- {1 \over 4 } \sum \Biggl[
{{J+L+M} \choose {R+S+Q-P-K}}I^N_{CD}-I^N_{DC} {{J+L+M} \choose
{R+S+Q-K-P-N}}\Biggr]
\]
\[
\times {{R+S+Q-K} \choose R}{K \choose Q}
\int_{\Omega} I^K_{BA}D_RE^J_B(h)
\]
\[
\times\biggl( D_SE^P_AE^L_C(f)
D_{J-R-S+M+L+N+P+K}E^M_D(g)-(F \leftrightarrow G)\biggr) ,
\]
we obtain
\[
\{ \{ F,G \} ,H \} ={1 \over 4}\sum
\Biggl[ {{J+L+M} \choose {R+S-P-K}}{{R+S-K} \choose {R-K}}
I^N_{CD}I^K_{AB}
\]
\[
-{{J+L+M} \choose {R+S-K-P-N}}{{R+S-K} \choose {R-K}}
I^N_{DC}I^K_{AB}
\]
\[
-\sum_Q{{J+L+M} \choose {R+S-P-K+Q}}{{R+S-K+Q} \choose R}
{K \choose Q}I^N_{CD}I^K_{BA}
\]
\[
+\sum_Q{{J+L+M} \choose {R+S-P-K-N+Q}}{{R+S-K+Q} \choose R}{K \choose Q}
I^N_{DC}I^K_{BA}\Biggr]
\]
\[
\times \int_{\Omega}
D_RE^J_B(h)\Biggl( D_SE^P_AE^L_C(f)D_{J+M+L+P+N+K-R-S}E^M_D(g)-
(F \leftrightarrow G)\Biggr) .
\]
Adding the terms with cyclic permutations, we group  terms like
\[
D_SE^P_AE^L_C(f)\biggl( D_RE^J_B(h)D_{J+M+L+P+N+K-R-S}E^M_D(g)
-(H \leftrightarrow G)\biggr),
\]
and, according to Lemma 2.8 substitute
\[
D_SE^P_AE^L_C(f)=D_S\sum (-1)^{\vert T \vert}
{{L+T} \choose L}E^{P-T}_A{{\partial f} \over {\partial \phi_C^{(L+T)}}}
\]
\[
=\sum_{T,U}(-1)^{\vert U \vert +\vert P \vert}{U \choose {P-T}}
{{L+T} \choose L}D_{S+U-P+T}{{\partial^2f} \over {\partial \phi_A^{(U)}
\partial \phi_C^{(L+T)}}}.
\]
Then make a change $T \rightarrow T-L$
\[
{1 \over 4}\sum (-1)^{\vert U \vert +\vert P \vert}{U \choose {L+P-T}}
{T \choose L}\lbrack\cdots\rbrack D_{S+U+T-L-P}
{{\partial^2f} \over {\partial \phi_A^{(U)}
\partial \phi_C^{(T)}}}
\]
\[
\times \biggl( D_RE^J_B(h)D_{J+M+L+P+N+K-R-S}E^M_D(g)
-(H \leftrightarrow G)\biggr) ,
\]
and $S \rightarrow S+L+P+N+K$
\[
{1 \over 4}\sum (-1)^{\vert U \vert +\vert P \vert}{T \choose L}
{U \choose {L+P-T}}
\]
\[
\times\Biggl[
{{J+L+M} \choose {R+S+L+N}}{{R+S+L+P+N}\choose {R-K}}I^N_{CD}I^K_{AB}
\]
\[
-{{J+L+M}\choose {R+S+L}}{{R+S+L+P+N}\choose {R-K}}I^N_{DC}I^K_{AB}
\]
\[
- \sum_Q{{J+L+M}\choose {R+Q+S+L+N}}{{R+S+Q+P+L+N}\choose R}{K \choose Q}
I^N_{CD}I^K_{BA}
\]
\[
+\sum_Q{{J+L+M}\choose {R+Q+S+L}}{{R+Q+S+L+P+N}\choose R}
{K \choose Q}I^N_{DC}I^K_{BA}\Biggr]
\]
\[
\times D_{S+U+T+N+K}
{{\partial^2f} \over {\partial \phi_A^{(U)}
\partial \phi_C^{(T)}}} \biggl( D_RE^J_B(h)D_{J+M-R-S}E^M_D(g)
-(H \leftrightarrow G)\biggr).
\]
So, we are able to estimate  sums over $P$
\[
\sum_P(-1)^{\vert P \vert}{U \choose {L+P-T}}{{R+S+L+P+N} \choose {R-K}}
\]
\[
=(-1)^{\vert L \vert +\vert U \vert + \vert T \vert}{{T+R+S+N} \choose
{R-K-U}},
\]
\[
\sum_P(-1)^{\vert P \vert}{U \choose {L+P-T}}{{R+S+L+P+N+Q} \choose
R}
\]
\[
=(-1)^{\vert L \vert +\vert U \vert + \vert T \vert}{{T+R+S+N+Q}
\choose {R-U}},
\]
and obtain
\[
{1 \over 4}\sum (-1)^{\vert L \vert +\vert T \vert}
{T \choose L}\Biggl[
{{J+L+M} \choose {R+S+L+N}}{{T+R+S+N}\choose {R-K-U}}I^N_{CD}I^K_{AB}
\]
\[
-{{J+L+M}\choose {R+S+L}}{{T+R+S+N}\choose {R-K-U}}I^N_{DC}I^K_{AB}
\]
\[
-\sum_Q{{J+L+M}\choose {R+Q+S+L+N}}{{T+R+S+Q+N}\choose {R-U}}{K \choose Q}
I^N_{CD}I^K_{BA}
\]
\[
+\sum_Q{{J+L+M}\choose {R+Q+S+L}}{{T+R+Q+S+N}\choose {R-U}}
{K \choose Q}I^N_{DC}I^K_{BA}
\Biggr]
\]
\[
\times D_{S+U+T+N+K}
{{\partial^2f} \over {\partial \phi_A^{(U)}
\partial \phi_C^{(T)}}}
 \biggl( D_RE^J_B(h)D_{J+M-R-S}E^M_D(g)
-(H \leftrightarrow G)\biggr).
\]
Summing over $L$
\[
\sum_L(-1)^{\vert L \vert}{T \choose L}{{L+J+M} \choose{L+R+S+N}}=
(-1)^{\vert T \vert}{{J+M}\choose {R+S+N+T}},
\]
\[
\sum_L(-1)^{\vert L \vert}{T \choose L}{{L+J+M} \choose{L+R+S}}=
(-1)^{\vert T \vert}{{J+M}\choose {R+S+T}},
\]
\[
\sum_L(-1)^{\vert L \vert}{T \choose L}{{L+J+M} \choose{L+R+S+N+Q}}=
(-1)^{\vert T \vert}{{J+M}\choose {R+S+T+N+Q}},
\]
\[
\sum_L(-1)^{\vert L \vert}{T \choose L}{{L+J+M} \choose{L+R+S+Q}}=
(-1)^{\vert T \vert}{{J+M}\choose {R+S+T+Q}},
\]
we get
\[
{1 \over 4}\sum\Biggl[
{{J+M} \choose {R+S+N+T}}{{T+R+S+N}\choose {R-K-U}}I^N_{CD}I^K_{AB}
\]
\[
-{{J+M}\choose {R+S+T}}{{T+R+S+N}\choose {R-K-U}}I^N_{DC}I^K_{AB}
\]
\[
-\sum_Q{{J+M}\choose {R+Q+S+T+N}}{{T+R+S+Q+N}\choose {R-U}}{K \choose Q}
I^N_{CD}I^K_{BA}
\]
\[
+\sum_Q{{J+M}\choose {R+Q+S+T}}{{T+R+Q+S+N}\choose {R-U}}{K \choose Q}I^N_{DC}
I^K_{BA}
\Biggr]
\]
\begin{equation}
\times D_{S+U+T+N+K}{{\partial^2f} \over {\partial \phi_A^{(U)}
\partial \phi_C^{(T)}}}\Biggl( D_RE^J_B(h)D_{J+M-R-S}E^M_D(g)
-(H \leftrightarrow G)\Biggr) .\label{eq:fin}
\end{equation}
Let us make change of indices $S \rightarrow -S-R+J+M$.
We can sum over $Q$ in the third term of the square brackets
\[
\sum_Q{{J+M}\choose {J+M-S+N+Q+T}}{{J+M-S+N+Q+T}\choose {R-U}}
{K \choose Q}
\]
\[
={{J+M}\choose {R-U}}{{J+M+K+U-R}\choose {S-N-T}}.
\]
Then after  interchanging $R \leftrightarrow S$, $J \leftrightarrow M$,
$B \leftrightarrow D$, $A \leftrightarrow C$, $N \leftrightarrow K$
and $U \leftrightarrow T$
we see that the first term in the square brackets stands as itself,
the new second term is equal to the old third and vice versa. The fourth
term transforms into itself\footnote{We are able only to verify this fact
by computer simulation.}:
\[
\sum_Q{{J+M} \choose {J+M-S+Q+T}}{{J+M-S+Q+T+N}\choose{R-U}}{K \choose Q}
\]
\[
= \sum_Q{{J+M} \choose {J+M-R+Q+U}}{{J+M-R+Q+U+K}\choose{S-T}}{N \choose Q}.
\]
Evidently, the round bracket in  Eq.(\ref{eq:fin}) changes its sign,
so the expression  is zero and the proof is completed.

\section{Conclusion}

It is clear that
the above results can be applied to the field theory  on manifolds with a
 boundary simply by postulating the Rule 4.2 and independently of
any reasoning about characteristic functions. The new Poisson structure
permits to consider dynamical problems in which boundary values of
hamiltonian variables are treated on equal footing with their internal values.
The dynamics of field variables on the boundary is determined  by both  volume
and  surface parts of the hamiltonian. Any choice of boundary conditions is in
fact a constraint in the phase space and should be treated along with standard
procedure of searching for secondary and higher constraints. These boundary
conditions do not interfere with the dynamical equations inside the domain
until we start solving elliptic type constraints, such as the Gauss law in
gauge theories. Then the nonlocal dependence appears, including the dependence
of surface variables, and surface part of the hamiltonian begins to  influence
the equations of internal variables (``divergencies cease to be
divergencies'' in terminology of Arnowitt, Deser and Misner \cite[p.434]{adm}
).

\vspace{12pt}

{\large\bf Acknowledgements}

The author is very grateful to S.N. Storchak for valuable comments
on the previous paper \cite{s2} and  recommending
book \cite{olv} without reading which this work could never been
done.  He is also  indebted to S.N. Talalov for bringing  papers \cite{app}
to the author's attention. It is a pleasure to thank A.V. Razumov for fruitful
comment on the role of characteristic function differentiating, E.G. Timoshenko
for useful discussions, V.F.Edneral and S.N.Sokolov for help in REDUCE
calculations.  The author is grateful to H.Nicolai for sending a copy of notes
of his lectures \cite{nic}  on Ashtekar's variables given at Winter school
in Karpac, which were important for understanding the unavoidability of
modifying Poisson brackets.  The completion of this paper was strongly
stimulated by support of A.I. Alekseev, S.I. Bitioukov, A.V. Kisselev, K.G.
Klimenko, G.L.  Rcheulishvili, A.P.Samokhin, S.M. Troshin  and C.Vosgien.

\hfill

\newpage

{\large \bf Appendix}

Here we list different forms in which the new Poisson brackets
can be written for the local case (\ref{eq:loc}):

1) through the full ({\it not standard})
variational derivatives defined by the formula (\ref{eq:varfull}) and
with account for Rule 5.4:
\[
\{ F,G\} =\sum\int\int {{\delta F} \over {\delta \phi_A(x)}}
\{ \phi_A(x),\phi_B(y) \} {{\delta G} \over {\delta \phi_B(y)}},
\]

2) through higher eulerian operators (\ref{eq:he}):
\[
{1 \over 2}\sum\int_{\Omega} D_{P+Q}\biggl( E^P_A(f)\hat I_{AB}E^Q_B(g)
 -E^P_A(g)\hat I_{AB}E^Q_B(f) \biggr) ,
\]
where
\[
\hat I_{AB}=\sum_NI^N_{AB}D_N,
\]

3) through Fr\'echet derivatives (\ref{eq:fr}):
\[
\{ F,G \} ={1 \over 2}
\sum\int_{\Omega} Tr(D_{f_A}\hat I_{AB}D_{g_B}- D_{g_A}\hat I
_{AB}D_{f_B}),
\]

4) through some matrix notations :
\[
\{ F,G \} ={1 \over 2}\sum\int_{\Omega}
 \Biggl( \langle\nabla f \cdot C \nabla g \rangle
- \langle\nabla g \cdot C \nabla
f\rangle \Biggr),
 \]
 defined below
 \[
 \nabla f=D_L{{\partial
 f} \over {\partial \phi_A^{(J)}}}, \qquad \nabla g=D_M{{\partial g} \over
{\partial \phi_B^{(K)}}},
\]
\[
 C_{JK,LM,AB}={J \choose L}{K \choose M}
D_{J+K-L-M}\hat I_{AB}.
\]

\end{document}